\documentstyle[preprint,floats,prl,aps,epsfig]{revtex}
\input psfig.sty

\begin{document}

\preprint{\tighten\vbox{\hbox{\hfil CLNS 00/1681}
                        \hbox{\hfil CLEO 00-13}
}}

\title{
First Observation of the $\Sigma_c^{*+}$ Baryon and a New Measurement 
of the $\Sigma_c^+$ Mass}

\author{CLEO COllaboration }
\date{\today}
\maketitle
\tighten
\begin{abstract}

Using data recorded with the CLEO II and CLEO II.V detector
configurations at the Cornell Electron Storage Ring, we 
report the first observation and mass measurement of the $\Sigma_c^{*+}$ 
charmed baryon, and an updated measurement of the mass of the $\Sigma_c^+$
baryon.
We find
$M(\Sigma_c^{*+})-M(\Lambda_c^+)=(231.0\pm1.1\pm2.0)\ {\rm MeV}$,
and $M(\Sigma_c^{+})-M(\Lambda_c^+)=(166.4\pm0.2\pm0.3)\ {\rm MeV}$, 
where the errors are statistical and systematic respectively.

\end{abstract}
\newpage
{
\renewcommand{\thefootnote}{\fnsymbol{footnote}}
\begin{center}
R.~Ammar,$^{1}$ A.~Bean,$^{1}$ D.~Besson,$^{1}$ R.~Davis,$^{1}$
N.~Kwak,$^{1}$ X.~Zhao,$^{1}$
S.~Anderson,$^{2}$ V.~V.~Frolov,$^{2}$ Y.~Kubota,$^{2}$
S.~J.~Lee,$^{2}$
% ...
\end{center}
 
\small
\begin{center}
$^{1}${University of Kansas, Lawrence, Kansas 66045}\\
$^{2}${University of Minnesota, Minneapolis, Minnesota 55455}\\
% ...
\end{center}

\setcounter{footnote}{0}

\newpage

The $\Sigma_c$ states consist of a charmed quark and two light 
($u$ or $d$) quarks, in an isospin one
configuration. 
The $J^P$=${1\over{2}}^+$ $\Sigma_c^{0}$ and $\Sigma_c^{++}$ have been 
observed for
many years. Their isospin partner, the $\Sigma_c^+$, is more difficult to 
detect 
as it decays to the $\Lambda_c^+$ with the 
emission of a neutral, as opposed to 
charged, pion. Neutral pion detection
is typically prone to higher backgrounds and poorer momentum resolution
than charged pion detection.
The $\Sigma_c^+$ was reported 
in one event in 1980\cite{CRAPP}, 
and then in a peak of 111 events by the CLEO collaboration 
in 1993\cite{SC+}. 
This analysis updates the earlier CLEO measurement with a much 
larger data sample. 
This permits a more accurate comparison of the isospin 
splitting of the $\Sigma_c$ states.

The $J^P={3\over{2}}^+$ $\Sigma_c^*$ states are more difficult to observe than
the $J^P={1\over{2}}^+$ states because of the larger natural width, which 
leads to a poorer signal to noise ratio. 
The $\Sigma_c^{*++}$ and $\Sigma_c^{*0}$ have now been identified in 
$\Lambda_c^+\pi^{\pm}$
final states, and their masses and widths measured\cite{SCS}. 
This analysis shows the first observation of their 
isospin partner, the
$\Sigma_c^{*+}$, observed by its decay to $\Lambda_c^+\pi^0$.  
This observation completes the spectroscopy of the seven L=0 $\Lambda_c$ and
$\Sigma_c$ baryons predicted by the quark model.

The data presented here 
were taken with the CLEO II and CLEO II.V detector configurations 
operating at the Cornell 
Electron Storage Ring (CESR).
The data sample used in this analysis corresponds to
an integrated luminosity of 13.7 $fb^{-1}$ taken on the $\Upsilon(4S)$ 
resonance and in the continuum at energies just below 
the $\Upsilon(4S)$.
Of this data, 4.7 $fb^{-1}$ was taken with the CLEO II configuration \cite{KUB}.
We detected charged tracks with a cylindrical drift chamber system inside
a 1.4T solenoidal magnet, and we detected photons using an electromagnetic
calorimeter consisting of 7800 cesium iodide crystals.
The remainder of the data was taken with the CLEO II.V 
configuration\cite{HILL}, which has
upgraded charged particle measurement capabilities, but the same same
cesium iodide array to observe photons.

In order to obtain large statistics we reconstructed the $\Lambda_c^+$
baryons using 15 different decay modes
\footnote{Charge conjugate modes are implicit throughout.}. 
Measurements of the branching
fractions into these modes have previously been presented by the CLEO
collaboration\cite{LAMC}, and the general procedures for finding
those decay modes can be found in those references.
For this search and data set, the exact analysis used has been optimized
for high efficiency and low background.
Briefly, particle identification of $p,K^-$, and $\pi$ candidates was performed
using specific ionization measurements in the drift chamber,
and, when present, time-of-flight measurements. Hyperons were found by
detecting their decay points separated from the main event vertex.

We reduce the combinatorial background, which is highest for
charmed baryon candidates with low momentum, by applying a cut on
$x_p$, where $x_p=p/p_{max}$, $p$ is the momentum
of the charmed baryon candidate, $p_{max}=\sqrt{E^2_{beam}-M^2},$ and $E_{beam}$ is the
beam energy, and $M$ is the reconstructed mass of the candidate.
 Using a cut of $x_p > 0.5$
(charmed baryons produced from decays of $B$ mesons 
near the $B\overline{B}$ threshold
are
kinematically limited to $x_p < 0.4$),
we fit the invariant mass distributions for these modes to a sum
of a Gaussian signal and a low-order polynomial background.
Combinations within $1.6 \sigma$ of the mass of the
$\Lambda_c^+$ in each decay mode are taken as $\Lambda_c^+$ candidates,
where the resolution of each decay mode is taken from a
Monte Carlo simulation
(for the CLEO II and CLEO II.V datasets separately). In this $x_p$ region,
we find a total yield of $\Lambda_c^+$ signal
of $\approx$ 58,000 combinations, and a signal-to-background ratio
$\approx 1:1.2$.

Photons were detected by their energy deposition in the crystal calorimeter. 
Each photon 
candidate was required to be well isolated from charged particles, 
and to have an energy profile
consistent with being due to a single photon. To ensure good 
signal to noise ratio, the transition $\pi^0$ candidates were made from the 
combination of 
two photons each from the central part of the detector ($\theta < 0.7)$, 
which has the best 
energy resolution. 
The calculated invariant 
mass of the photon pair was required to be within 2.5 standard deviations 
of the known $\pi^0$ mass, and the momentum of the $\pi^0$ candidate
was required to be 
greater than 150 MeV/c. This momentum cut was optimized to maximize the signal 
to noise ratio 
of a resonance in the expected $\Sigma_c^{*+}$ mass range using a Monte Carlo 
simulation.
The $\pi^0$ candidates were then kinematically fit 
to the $\pi^0$ mass, a procedure that improves the mass resolution of the 
$\Sigma_c^{*+}$ by around twenty percent.

The $\Lambda_c^+$ candidates were 
combined with each $\pi^0$ candidate in the event 
and the mass difference 
$M(\Lambda_c^+\pi^0)-M(\Lambda_c^+)$ was calculated.
Our requirement on the fractional momentum, 
$x_p > 0.6$, is placed on the $\Lambda_c^+\pi^0$ combination, not 
on the $\Lambda_c^+$ itself. Given the energetics of the decays to 
$\Lambda_c^+\pi^0$, such a criterion corresponds roughly
to $x_p > 0.5$ for the $\Lambda_c^+$ daughters. 
The mass difference spectrum, shown in Figure 1, shows two clear 
peaks. The first, near 167 $ \rm{MeV}$, is due to $\Sigma_c^+$ decays. 
The second, near $230\ \rm{MeV}$, we identify as the $\Sigma_c^{*+}$.
If we fit this distribution to the sum of a third-order 
Chebychev polynomial distribution 
and two Gaussian signals, we obtain a yield of $661^{+63}_{-60}$ events and 
a width of $\sigma =\ $  
($2.84^{+0.31}_{-0.28})\ $MeV for the $\Sigma_c^+$, and a yield of 
$(327^{+78}_{-73})$ events 
and  $\sigma$ = $(5.6\pm1.4)\ $MeV for the second peak. 
The widths of these Gaussian signals are greater than the 
detector resolution, calculated from a GEANT-based\cite{GEANT}
Monte Carlo simulation program, of 
1.90 and 3.55 MeV, respectively, in the relevant mass regions, 
indicating the likelihood that the particles 
have non-negligible intrinsic widths. If we fit the distribution instead 
to a sum of 
two p-wave Breit-Wigner functions convoluted 
with Gaussian resolution functions, we 
obtain values of the intrinsic width, 
$\Gamma$, of $(3.1^{+0.9}_{-0.8})\ $ MeV, and
$(7^{+6}_{-5})\ $ MeV respectively, for which the errors are statistical only. 
The pole masses obtained from this fit are 
$M(\Sigma_c^+)-M(\Lambda_c^+)=(166.44\pm0.24)$ MeV and
$M(\Sigma_c^{*+})-M(\Lambda_c^+) $= $(231.0\pm 1.1) \ $ MeV, 
where again the quoted errors are from the
statistical errors in the fit. It is this second fit, which has a $\chi^2$ of 73.3 for 93
degrees of freedom, which is shown in Figure 1. If the $\Sigma_c^{*+}$ signal 
were not included in the fit, it would have a $\chi^2$ of 123 for 96 
degrees of freedom.
To obtain an estimate of the relative cross sections for $\Lambda_c^+$, 
$\Sigma_c^+$ and $\Sigma_c^{*+}$ baryons, we find the yield each of the three
states with an $x_p$ cut on each candidate of 0.6.
After correcting for the efficiency
of the transition $\pi^0$, we find the ratio  $N(\Sigma_c^+$):$N(\Lambda_c^+$)=
$0.116^{+0.016}_{-0.014}$$\pm$$0.022$ and 
$N(\Sigma_c^{*+})$:$N(\Lambda_c^+)$=$0.043^{+0.016}_{-0.012}$$\pm$$0.007$, where the 
errors are statistical and systematic respectively. The systematic uncertainty includes 
the uncertainty in the $\pi^0$ reconstruction efficiency and differences in the
yield obtained with different signal shapes.
We note that we are not calculating the production 
ratios of these states, as we are unable to measure their full momentum spectra. 

We have considered many different possible sources of systematic 
uncertainty in the 
measurements of the masses and widths of these resonances. We have checked the 
consistency of the results obtained with each of the two detector configurations 
separately, 
as well as with a variety of different background and signal shapes, different
criteria on the $\pi^0$ momenta, and different $\Lambda_c^+$ decay modes. 
We find the dominating 
systematic uncertainties in the mass measurement of the $\Sigma_c^+$ 
to be due to 
signal shape (0.2 MeV) and the uncertainty in the $\pi^0$ momentum measurement 
(0.2 MeV). These combine to give a total systematic uncertainty in the measurement
of $M(\Sigma_c^+)$ of 0.3 MeV. 
In the case of the
$\Sigma_c^{*+}$, the mass measurement is sensitive to 
both the shape of the signal
and also to the shape of the background function used, and we estimate a 
total systematic
uncertainty of 2 MeV in the measurement of the pole mass. 
Although the intrinsic width measurement of the $\Sigma_c^+$ is 
statistically nearly four standard deviations from 
0, there should also be added a systematic uncertainty which we estimate to be
0.8 MeV, due mostly to uncertainties in the energy resolution of the 
transition pion.
The combination of statistical and systematic uncertainties lead us to 
set an upper limit of 4.6 MeV (at the 90\% confidence level)
on $\Gamma(\Sigma_c^+)$. The width of the $\Sigma_c^{*+}$ is particularly 
sensitive to the 
parameterization of the background shape, and we estimate a systematic uncertainty of 
5 MeV in the measurement of $\Gamma(\Sigma_c^{*+})$ mostly from this source. This,
combined with the statistical error, leads to a 90\% confidence level limit of
$\Gamma<17\ $ MeV.

Our result for the mass of the $\Sigma_c^+$ is rather lower than the previous CLEO measurement
which was based upon a small subset of these data, 
and lower than the measured masses of the $\Sigma_c^{++}$ and $\Sigma_c^0$, for which 
more experimental data is available\cite{PDG}. This is consistent with the theoretical expectations
for this isospin splitting \cite{FRANK}. The mass of the $\Sigma_c^{*+}$ is also lower
than that of its isospin partners, but the experimental errors are too 
large for this splitting
to be significant.

In conclusion, we have made a new measurement of the mass of the $\Sigma_c^+$
and find $M(\Sigma_c^+)-M(\Lambda_c^+)$ = $ (166.4\pm0.2\pm0.3)$ MeV. We report the first
observation of the $\Sigma_c^{*+}$ and find 
$M(\Sigma_c^{*+})-M(\Lambda_c^+)$ = $(231.0\pm1.1\pm2.0)$
MeV. These measurements are consistent with expectations based upon the previously observed
isospin partners of these two particles.

\smallskip

We gratefully acknowledge the effort of the CESR staff in providing us with
excellent luminosity and running conditions.
This work was supported by 
the National Science Foundation,
the U.S. Department of Energy,
the Research Corporation,
the Natural Sciences and Engineering Research Council of Canada, 
the A.P. Sloan Foundation, 
the Swiss National Science Foundation, 
the Texas Advanced Research Program,
and the Alexander von Humboldt Stiftung.

\begin{figure}[htb]
\noindent
\psfig{bbllx=60pt,bblly=100pt,bburx=440pt,bbury=770pt,
file=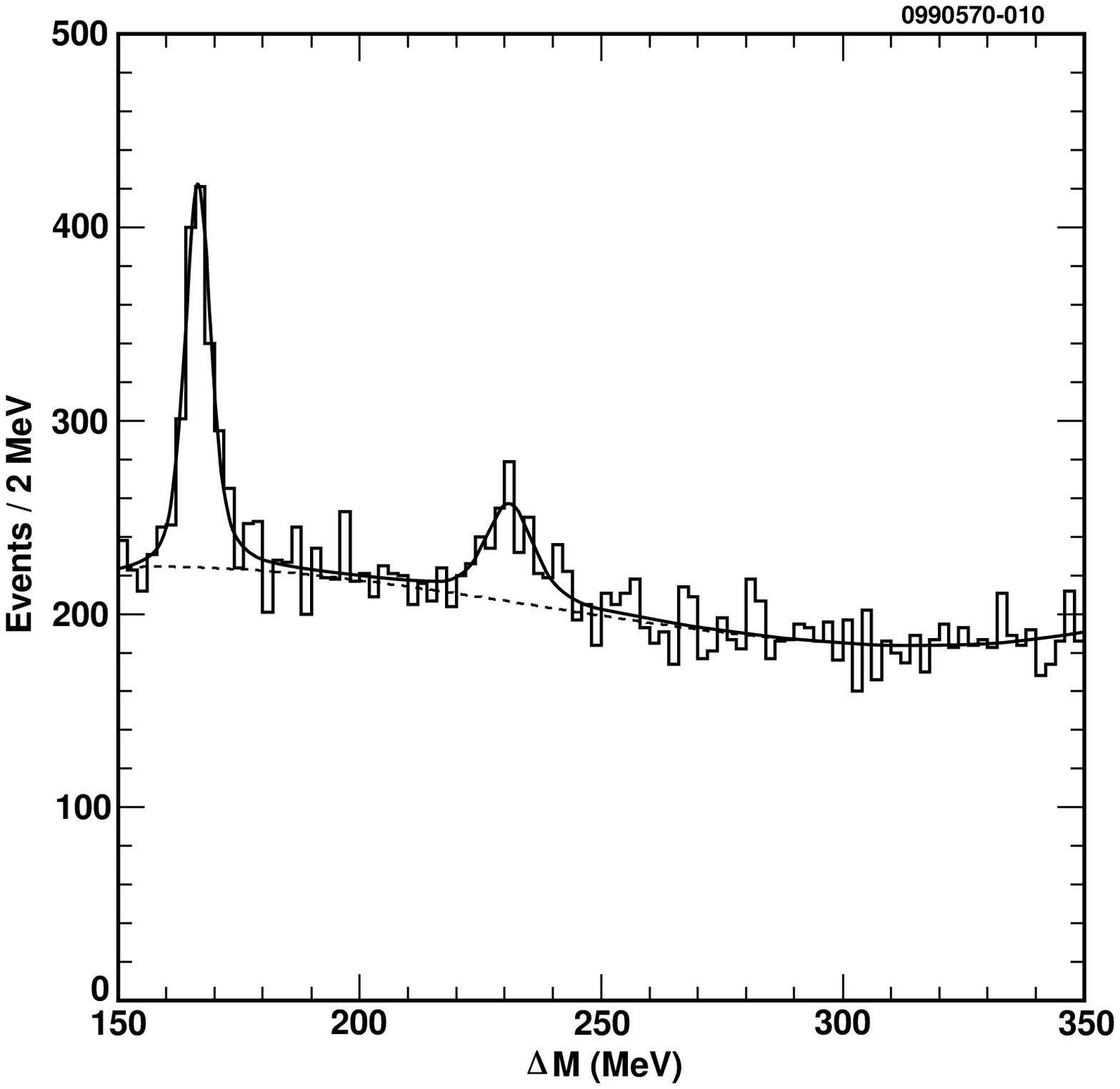,width=4.5in}
\caption[]{Mass difference spectrum, $M(\Lambda_c^+\pi^0)-M(\Lambda_c^+)$.
The solid line fit is to a third-order polynomial background shape and two p-wave
Breit-Wigner functions smeared by Gaussian resolution functions for the
two signal shapes. The dashed line shows the background function.}
\end{figure}
\vfill

\narrowtext
\end{document}